\documentstyle[12pt,epsf]{article}
\newlength{\extraspace}
\newlength{\extraspaces}
\catcode`\@=11
\def\numberbysection{\@addtoreset{equation}{section}
\def\theequation{\arabic{section}.\arabic{equation}}}
\newcommand{\be}{\begin{equation}
\addtolength{\abovedisplayskip}{\extraspaces}
\addtolength{\belowdisplayskip}{\extraspaces}
\addtolength{\abovedisplayshortskip}{\extraspace}
\addtolength{\belowdisplayshortskip}{\extraspace}}
\newcommand{\ee}{\end{equation}}
\newcommand{\ba}{\begin{eqnarray}
\addtolength{\abovedisplayskip}{\extraspaces}
\addtolength{\belowdisplayskip}{\extraspaces}
\addtolength{\abovedisplayshortskip}{\extraspace}
\addtolength{\belowdisplayshortskip}{\extraspace}}
\newcommand{\ea}{\end{eqnarray}}
\newcommand{\newsection}[1]{
\vspace{7mm}
\pagebreak[3]
\addtocounter{section}{1}
\setcounter{equation}{0}
\setcounter{subsection}{0}
\setcounter{footnote}{0}
\begin{center}
{\large {\bf \thesection. #1}}
\end{center}
\nopagebreak
\medskip
\nopagebreak
\hspace{3mm}}

\newcommand{\Tr}{{\rm Tr}}

\newcommand{\one}{{\bf 1}}
\newcommand{\bm}[1]{\mbox{\boldmath$ #1 $}}
\def\barr{\left(\!\!\begin{array}}
\def\earr{\end{array}\!\!\right)}
\def\(({\left(}
\def\)){\right)}

\begin{document}
\addtolength{\baselineskip}{.7mm}
\thispagestyle{empty}
\begin{flushright}
TIT/HEP--264 \\
{\tt hep-ph/9503383} \\
March, 1995
\end{flushright}
\vspace{2mm}
\begin{center}
{\Large{\bf Neutron Electric Dipole Moment in the Minimal
Supersymmetric Standard Model}} \\[5mm]
{\sc Tomoyuki Inui}
\footnote{JSPS Fellow, {\tt e-mail: kankun@th.phys.titech.ac.jp}},
{\sc Yukihiro Mimura}
\footnote{\tt e-mail: mim@th.phys.titech.ac.jp},
\\[4mm]
{\sc Norisuke Sakai}
\footnote{\tt e-mail: nsakai@th.phys.titech.ac.jp},
\ \
and
\ \
{\sc Tomoharu Sasaki}
\footnote{\tt e-mail: tsasaki@th.phys.titech.ac.jp} \\[5mm]
{\it Department of Physics,
Tokyo Institute of Technology, \\[2mm]
Oh-okayama, Meguro, Tokyo 152, Japan} \\[10mm]
{\bf Abstract}\\[5mm]
{\parbox{13cm}{\hspace{5mm}
Neutron electric dipole moment (EDM) due to single quark EDM and to
the transition EDM is calculated in the minimal
supersymmetric standard model.
Assuming that the Cabibbo-Kobayashi-Maskawa matrix at the
grand unification scale is the only source of CP violation,
complex phases are induced in parameters of soft supersymmetry
breaking at low energies.
Chargino one-loop diagram is found to give the dominant contribution
of the order of $10^{-27}\sim 10^{-29}\:e\cdot$cm for quark EDM,
assuming the light chargino mass and the universal scalar mass to be
$50$ GeV and $100$ GeV, respectively.
Therefore the neutron EDM in this class of model
is difficult to measure experimentally.
Gluino one-loop diagram also contributes due to the flavor changing
gluino coupling.
The transition EDM is found to give dominant contributions for certain
parameter regions.
}}

\end{center}
\vfill
\newpage
\setcounter{section}{0}
\setcounter{equation}{0}
\numberbysection




\newsection{Introduction}

The supersymmetric theories now stand as the most promising candidate
for the unified theory beyond the standard model \cite{Nilles}.
The supersymmetry helps to resolve the gauge hierarchy problem
\cite{DG}.
Moreover, the accurate data favor remarkably the supersymmetric
grand unified theory (GUT) over the nonsupersymmetric
theory \cite{Amaldi}.
This fact has fulfilled the promise that accurate measurements of
coupling constant strengths at low energies can distinguish various
alternative candidates for the grand unified theories by extrapolating
the renormalization group trajectories to higher energies.

Among many problems in particle physics, the violation of $CP$
invariance is one of the phenomena that are least understood.
The primary reason for this unsatisfactory situation is that
the experimental verification of the
$CP$ violation is so far limited to neutral Kaon dacays into two pions.
We expect to obtain more experimental informations on the $CP$
violation from the B-factory soon.
The $CP$ violation is not only important as a fundamental
symmetry property, but also needed to explain
the cosmological baryon asymmetry of our universe \cite{Sakharov}.
Therefore
it is most desirable to have additional experimental
informations on the $CP$ violation.
Apart from the forthcoming experiment with the B-factory, we
have one more promising observable for the $CP$ violation:
the electric dipole moment (EDM), in particular those of neutron and
electron \cite{Barr}.
Since we can hope for further improvements of experimental
precision, especially for that of neutron, we expect that
the EDM will provide
a precious clue of the $CP$ violation.
The $CP$ violation in the minimal standard model
arises solely from the Cabibbo-Kobayashi-Maskawa matrix of the Yukawa
coupling constants of the Higgs field.
Therefore progress in the study of the $CP$ violation provides
important informations on the Higgs field which is most elusive in
the standard model.

In supersymmetric models, we have more possibilities for complex
parameters beside the Cabibbo-Kobayashi-Maskawa matrix,
even if we have only the minimal particle content of the
supersymmetric standard model.
These complex parameters
become additional sources of the $CP$ violation.
Among the parameters of the supersymmetric models, those associated
with the soft breaking of supersymmetry are least understood.
The early studies of the EDM in supersymmetric models
have revealed that generic complex parameters for the soft breaking
give too large EDM unless superpartners in the loop
are heavier than at least a few TeV \cite{Kizukuri}.
Although this type of models are not excluded, it is perhaps more
attractive and natural if we can control the phases of the soft
breaking parameters so that superpartner masses of the order of the
electroweak scale are naturally allowed.
There have been many studies on this issue
\cite{Buchmuller,Weinberg}.

In the most popular model, the supergravity with the hidden sector
provides a definite pattern of the soft breaking of
supersymmetry \cite{Nilles,NAC}.
If we assume a simple model for the hidden sector and the supergravity
couplings, we obtain that these parameters are real at the grand
unification scale or the Planck scale.
However, the soft breaking parameters that are really manifest at low
energies will become complex since nonvanishing phases will be induced
by the renormalization group flow involving the
Cabibbo-Kobayashi-Maskawa matrix \cite{BV,Barr}.
Moreover, flavor-changing couplings of gluino are also induced
\cite{AGN}.
This radiative effect becomes important when the Yukawa couplings
are large.
It is worth examining the $CP$ violation due to this radiatively
induced phases of the soft breaking parameters, since it is now
certain that the top quark is quite heavy \cite{CDF} and requires a
large Yukawa coupling.

More recently, there have been a number of studies on the neutron EDM
in the supersymmetric models \cite{DIHA}
or in two Higgs doublet models \cite{HKMTW}.

In the nonsupersymmetric minimal standard model, it has been proposed
that the transition quark EDM can be more important
than the single quark EDM to explain the neutron
EDM \cite{Morel,GAV,GAVpenguin}.

The purpose of this paper is to examine the neutron electric dipole
moment in the minimal supersymmetric standard model.
We assume that there are complex parameters only in the
Cabibbo-Kobayashi-Maskawa matrix at the grand unification scale.
We examine the effect of the radiatively induced phases of the soft
supersymmetry breaking parameters on the neutron EDM.
In performing the renormalization group analysis, we have taken
account of the effect of gaugino masses together with the universal
scalar masses.
We shall also consider the transition quark EDM
in the supersymmetric models.
We find that the single quark EDM is of the order of
$10^{-27} \sim 10^{-29}\:e\cdot$cm for the light chargino mass to be
50 GeV.
We also find that the transition EDM is
of the order of $10^{-25} \sim 10^{-27}\:e\cdot$cm and hence
contributes the neutron EDM of the order of
$10^{-29} \sim 10^{-31}\:e\cdot$cm, if we take account of
the large QCD enhancement due to the penguin diagrams.
In both cases, the neutron EDM in this class of
models is too small to be detected in forthcoming experiments.
\par
In sect.\ 2, we introduce soft breaking parameters of
supersymmetry and scalar particle mass matrices.
In sect.\ 3, we analyze the single quark EDM.
There are two classes of contributions: the chargino loop
and the gluino loop.
In sect.\ 4, we examine the transition EDM.
Appendix is devoted to describe our results of the renormalization
group equations and our inputs.


\newsection{Soft breaking parameters of supersymmetry}

 We consider the minimal supersymmetric standard model (MSSM)
which contains left-chiral supermultiplets for three generations
of quarks ($\bm{U}^c$, $\bm{D}^c$, $\bm{Q}$)
and leptons ($\bm{E}^c$, $\bm{L}$), gauge bosons of
the SU$(3)_{\rm c}\times$SU$(2)_{\rm L}\times$U$(1)_Y$ and two Higgs
doublets ($H_u$, $H_d$).
Boldface letters such as $\bm{Q}$ denote vectors in generation
indices, and the suffix $c$ denotes the antiparticle.
The supergravity with the hidden sector provides the pattern of the
soft breaking of supersymmetry \cite{Nilles,NAC}.
In simple models of this type, the soft breaking parameters are
real and universal at the grand unification (GUT) scale or the Planck
scale,
and a complex phase appears only in the Cabibbo-Kobayashi-Maskawa
matrix at the scale. Renormalization group
running induces complex phases in the soft breaking parameters at the
lower scale.

One can write superpotential of MSSM as follows,
\be
 -W\equiv\bm U^{cT}\bm Y_u H_u\cdot\bm Q-\bm D^{cT}\bm Y_d H_d\cdot\bm
  Q-\bm E^{cT}\bm Y_e H_d\cdot\bm L+\mu H_u\cdot H_d,
\ee
where boldface letters $\bm{Y}$ are Yukawa couplings as a matrix
in generation indices.
The inner product of SU(2) indices is abbreviated by the $\cdot$ as
$H_u\cdot H_d \equiv (i\sigma_2H_u)^TH_d$.

Soft supersymmetry breaking is given by the following terms
in the Lagrangian:
\begin{enumerate}
\item scalar mass terms,
\be
 - \sum_{i,j}m_{ij}^2 \phi_i^* \phi_j
\ee
The $\phi_i$'s are all the scalar particles and $m_{ij}^2$ is a
hermitian matrix.  In the supergravity-induced models, this will be
universal at the GUT scale, i.e., $m_{ij}^2=m^2\delta_{ij}$ at the
GUT scale. It runs by renormalization group flow.

\item $A$ terms (trilinear scalar couplings),
\be
 -\tilde{\bm{u}}_R^{\dagger} \bm{A}_u h_u \cdot
\tilde{\bm{q}}_L +
 \tilde{\bm{d}}_R^{\dagger} \bm{A}_d h_d \cdot
\tilde{\bm{q}}_L +
 \tilde{\bm{e}}_R^{\dagger} \bm{A}_e h_d \cdot
\tilde{\bm{\ell}}_L +
{\rm H.c.}
\ee
In the supergravity-induced models, $\bm A$'s are proportional
to Yukawa couplings at the GUT scale, $\bm A=A\bm Y$.
The $\bm A$ terms run by renormalization group flow.

\item $B$ term,

\be
 B \mu h_u \cdot h_d + {\rm H.c.}
\ee

\item gaugino mass terms,
\be -\sum_{i=1}^3M_i\overline{\lambda_{iR}}\lambda_{iL}+{\rm H.c.} \ee
If GUT is embedded in the supergravity-induced models, $M_i$
($i=1,2,3$) are universal,
i.e., $M_i=M$ at the GUT scale. Gaugino masses $M_i$ run by
renormalization group flow.

\end{enumerate}

We can write scalar-quark mass terms in the following form:
\be
 -\barr{cc}\tilde{\bm{u}}_L^{\dagger}&\tilde{\bm{u}}_R^{\dagger}\earr\:
    M_{\tilde{u}}^2
  \barr c\tilde{\bm{u}}_L \\
         \tilde{\bm{u}}_R \earr
 -\barr{cc}\tilde{\bm{d}}_L^{\dagger}&\tilde{\bm{d}}_R^{\dagger}\earr\:
    M_{\tilde{d}}^2
  \barr c\tilde{\bm{d}}_L \\
         \tilde{\bm{d}}_R \earr,
\ee
where $\tilde{\bm{u}}_{L,R}$ ($\tilde{\bm{d}}_{L,R}$) is a
u-type-scalar-quark
(d-type-scalar-quark) field column vector in generation indices.

Denoting the scalar-quark mass matrices at the GUT scale with
the suffix $0$, one can find
\ba
M_{\tilde u0}^2 &=&
\barr{cc}
m_{\tilde uL}^2 \one + \bm{M}_u^{\dagger} \bm{M}_u &
(A+\mu \cot \beta) \bm{M}_u^{\dagger} \\
 \bm{M}_u (A+\mu \cot \beta)  & m_{\tilde uR}^2 \one +
\bm{M}_u \bm{M}_u^{\dagger}
\earr, \\
M_{\tilde d0}^2 &=&
\barr{cc}
m_{\tilde dL}^2 \one + \bm{M}_d^{\dagger} \bm{M}_d &
(A+\mu \tan \beta) \bm{M}_d^{\dagger} \\
\bm{M}_d (A+\mu \tan \beta)  & m_{\tilde dR}^2 \one +
\bm{M}_d \bm{M}_d^{\dagger}
\earr,
\ea

where $\bm{M}_u$ and $\bm{M}_d$ are matrices in generation
indices for u-type and d-type quark masses respectively and
are given by Yukawa couplings $\bm{Y}$ and vacuum expectation values
$v$ of Higgs fields,
\ba
\bm M_u=\bm Y_u\frac{v_u}{\sqrt2},&&\bm M_d
=\bm Y_d\frac{v_d}{\sqrt2}.
\ea
The $\tan \beta$ is defined as $v_u / v_d$.
The generation independent part of the scalar-quark masses are
given by the universal mass and the contribution from the $D$-term
 \cite{MaRa}:
\ba
 m_{\tilde uL}^2&=&m^2-M_Z^{\ 2}(-\cos2\beta)
\left(\frac12-\frac23\sin^2\theta_W\right),\\
 m_{\tilde uR}^2&=&m^2-\frac23M_Z^{\ 2}(-\cos2\beta)\sin^2\theta_W,\\
 m_{\tilde dL}^2&=&m^2+M_Z^{\ 2}(-\cos2\beta)
\left(\frac12-\frac13\sin^2\theta_W\right),\\
 m_{\tilde dR}^2&=&m^2+\frac13M_Z^{\ 2}(-\cos2\beta)\sin^2\theta_W.
\ea

In writing the above formulas, we assumed for simplicity
a universal form of the
supersymmetry breaking, namely the universal scalar mass $m$ and
the trilinear scalar coupling with the universal parameter $A$.
Moreover, all the parameters are real
except the Yukawa coupling constants which appear in the quark
mass matrices at the GUT scale.
Therefore the Yukawa couplings or the Cabibbo-Kobayashi-Maskawa matrix
is the only source of $CP$ violation in this model at the GUT scale.

The renormalization group flow changes these scalar-quark mass
matrices at the lower scale \cite{IKKT}.
At the lower scale, these mass matrices can be written as follows:
\ba
 M_{\tilde u}^2 &=&
  \barr{cc}
m_{\tilde uL}^2\one+\bm{M}_u^\dagger\bm{M}_u+\bm{\delta m}_{\tilde
uL}^2
 &
    (\bm{A}_u+\mu \cot \beta\one) \bm{M}_u^{\dagger} \\
 \bm{M}_u(\bm{A}_u^\dagger+\mu\cot\beta\one) & m_{\tilde
  uR}^2\one+\bm{M}_u\bm{M}_u^\dagger+\bm{\delta m}_{\tilde uR}^2
  \earr,
\label{eqn:scumass}\\
 M_{\tilde d}^2 &=&
  \barr{cc}
m_{\tilde dL}^2\one+\bm{M}_d^\dagger\bm{M}_d+\bm{\delta m}_{\tilde
  dL}^2
&
    (\bm{A}_d+\mu \tan \beta\one) \bm{M}_d^{\dagger} \\
 \bm{M}_d(\bm{A}_d^\dagger+\mu\tan\beta\one) & m_{\tilde
 dR}^2\one+\bm{M}_d\bm{M}_d^\dagger+\bm{\delta m}_{\tilde dR}^2
  \earr,
\label{eqn:scdmass}
\ea
where
$\bm{\delta m}^2$'s are hermitian matrices defined from the soft SUSY
breaking scalar mass parameters which are determined from the
renormalization group flow and have off-diagonal elements at the
electroweak scale.
After the renormalization group running,
the parameter $A$'s become matrices in generation indices
and the left-left
(LL) and right-right (RR) blocks have off-diagonal elements.
The off-diagonal terms of the matrices $\bm A$'s
turn out to be important for the neutron EDM.  We show the
renormalization group
equations and their solutions for the matrices $\bm A$'s in Appendix
\ref{adx:rge}.

It is convenient to rotate
the scalar-quark wave functions by the same amount as to diagonalize
the mass matrices of the quarks themselves, although scalar-quarks are
not in mass eigenstates in this basis.
This basis of the scalar-quark wave function is usually called
super KM basis.
Denoting the wave functions in the super KM basis with a prime,
we obtain quarks and scalar-quarks as
\ba
\barr c\bm u'_L\\
         \tilde{\bm u}'_L\earr&=&\bm V_{uL}\barr c\bm u_L\\
                                                 \tilde{\bm
                 u}_L\earr,\\
   \barr c\bm d'_L\\
         \tilde{\bm d}'_L\earr&=&\bm V_{dL}\barr c\bm d_L\\
                                                 \tilde{\bm
                 d}_L\earr,\\
   \barr c\bm u'_R\\
         \tilde{\bm u}'_R\earr&=&\bm V_{uR}\barr c\bm u_R\\
                                                 \tilde{\bm
                 u}_R\earr,\\
   \barr c\bm d'_R\\
         \tilde{\bm d}'_R\earr&=&\bm V_{dR}\barr c\bm d_R\\
                                                 \tilde{\bm d}_R\earr,
\ea
where
the quark mass matrices are diagonalized in generation indices
with the matrices $V_{uL}$ and so on:
\ba
 \bm V_{uR}\bm M_u\bm V_{uL}^\dagger&=&\bm m_u,\label{eqn:rotatemu}\\
 \bm V_{dR}\bm M_d\bm V_{dL}^\dagger&=&\bm m_d.\label{eqn:rotatemd}
\ea
 The Cabibbo-Kobayashi-Maskawa Matrix is defined as
$\bm K=\bm V_{uL}\:\bm V_{dL}^\dagger$.
We use a parametrization and explicit values of $\bm K$ as given in
Eqs.\ (\ref{ckmmatrix}) -- (\ref{delta}) of Appendix \ref{adx:rge}.
We also rotate $\bm{\delta m}^2$'s and $\bm A$'s as follows:
\ba
 \bm{\delta m}_{\tilde fH}^{2\prime}&=&
    \bm V_{fH}\bm{\delta m}_{\tilde fH}^2\bm V_{fH}^\dagger,\\
 \bm A_f'          &=&\bm V_{fL}\bm A_f\bm V_{fL}^\dagger,
 \label{eqn:rotatea}
\ea
where $f=u,d$ and $H=L,R$.

Scalar-quark mass matrices in this basis are given by
\ba
 M_{\tilde u}^{\prime2}&=&\barr{cc}
                        \bm V_{uL}&0\\
                        0         &\bm V_{uR}
                        \earr     M_{\tilde u}^2\barr{cc}
                                       \bm V_{uL}^\dagger&0\\
                                           0 &\bm
                                            V_{uR}^\dagger
                                             \earr\nonumber\\
                       &=&\barr{cc}
 m_{\tilde uL}^2\one+\bm m_u^2+\bm{\delta m}_{\tilde uL}^2
                      &(\bm A_u+\mu\cot\beta\one)\bm m_u\\
                   \bm m_u(\bm A_u^\dagger+\mu\cot\beta\one)
                             &m_{\tilde uR}^2\one+\bm m_u^2
                          +\bm{\delta m}_{\tilde uR}^2
                          \earr,
\label{scalarmass}\\
 M_{\tilde d}^{\prime2}&=&\barr{cc}
                           \bm V_{dL}&0\\
                           0         &\bm V_{dR}
                          \earr            M_{\tilde d}^2\barr{cc}
                            \bm V_{dL}^\dagger&0\\
                                           0  &\bm V_{dR}^\dagger
                                                  \earr\nonumber\\
                       &=&\barr{cc}
     m_{\tilde dL}^2\one+\bm m_d^2+\bm{\delta m}_{\tilde dL}^2
                           &(\bm A_d+\mu\tan\beta\one)\bm m_d\\
                        \bm m_d(\bm A_d^\dagger+\mu\tan\beta\one)
       &m_{\tilde dR}^2\one+\bm m_d^2+\bm{\delta m}_{\tilde dR}^2
                          \earr.
\ea
Here we neglected the primes in the right hand sides.  Hereafter we
consider in this basis.  We show $\bm A$'s at the electroweak scale
in this basis which are solutions of renormalization group equations
in Appendix \ref{adx:rge}.  We give initial conditions for $\bm A$'s
at the GUT scale $M_{GUT}$ as in Eq.\ (\ref{eqn:ainitial}).  The $\bm
A$'s are diagonal at the GUT scale.  Then $\bm A_e$
is diagonal at all the scale, but $\bm A_u$ and $\bm A_d$ are not
diagonal at the lower scale.  It is convenient to separate $\bm A$'s
into two parts,
\be \bm A_f=\bm A_{Lf}+\bm A_{Mf}, \ee
where $f=u,d,e$.
The second term $\bm A_{Mf}$
is proportional to the universal gaugino mass $M$ and satisfies
the same renormalization group
equations (\ref{eqn:rgeau}) -- (\ref{eqn:rgeae})
as $\bm A_{f}$.
Their initial conditions at the GUT scale are
\be \bm A_{Mu}(0)=\bm A_{Md}(0)=\bm A_{Me}(0)=0. \ee
The first term $\bm A_{Lf}$ satisfies linear equations
(\ref{eqn:rgealu}) -- (\ref{eqn:rgeale})
which are obtained by deleting gaugino masses in
Eqs.\ (\ref{eqn:rgeau}) -- (\ref{eqn:rgeae}).
Our results of the renormalization group analysis are summarized in
the Appendix by giving the matrix $\bm A_f$ at low energies in
Eqs.\ (\ref{eqn:valuealu}) -- (\ref{eqn:valueame}).


\newsection{Single quark electric dipole moment}

The matrix element of the electromagnetic current $j^\mu(0)$ between
a Dirac fermion with momentum $\bm{p}$ and spin component $s$
can be written in terms of four independent form factors $F_i$
as follows:
\ba
\langle\bm{p}_f,s_f|j^\mu(0)|\bm{p}_i,s_i\rangle
 &\!\!\!= &\!\!\!\overline{u}(\bm{p}_f,s_f)
\left[\gamma^\mu F_1(q^2)+i\sigma^{\mu\nu}q_\nu\frac{F_2(q^2)}{2m}
     +\gamma_5\sigma^{\mu\nu}q_\nu\frac{F_3(q^2)}{2m}\right.
                                                        \nonumber\\
&\!\!\! &\!\!\!
\qquad\qquad
\quad
+\left.\((\frac{q^2}{2m}\gamma^\mu-q^\mu\))\gamma_5F_A(q^2)\right]
u(\bm{p}_i,s_i),
\ea
where $q=p_f-p_i$.
The electric dipole moment (EDM) $d$ of the spin-$\frac12$ particle
is given in
terms of the form factor $F_3(q^2)$ as
\be
 d=-\frac e{2m}F_3(0),
\ee
since a Dirac particle with spin vector $\bm s$ and with
the EDM $d$ interacts with
the weak external electric field $\bm E$ as
\be
 L_{\rm int}=2d\bm s\cdot\bm E.
\ee
This interaction violates the $CP$ invariance.

We calculate the neutron EDM in the MSSM.
Contrary to the minimal standard model without supersymmetry,
we have contributions already at one-loop.
There are two classes of contributions:

(a) chargino loop, as shown in Fig.~\ref{chrd}

(b) gluino loop, as shown in Fig.~\ref{glud}

Although the complex phase is assumed to be only in the
Cabibbo-Kobayashi-Maskawa matrix at the GUT scale,
the matrices $\bm{\delta m}_L^2$ and $\bm{A}$ in
the scalar-quark mass matrices also have complex phases at low
energies as discussed in the previous section.
We evaluate the contribution of these radiatively
induced phases to the EDM.

There are a few diagrams which
give significant contributions to EDM.
Especially, the diagram of Fig.~\ref{chrd} gives a dominant
contribution for EDM since it involves the Yukawa coupling constant
of the top quark most directly.
The EDM from the diagram in Fig.~\ref{chrd} can be written as
\be
d_d=\frac e{16\pi^2}(\tan\beta+\cot\beta)\frac2{v^2}m_d\,
{\rm Im}\!
\left[\bm K^\dagger\left\{F\!\((
\frac{M_{\tilde u}^{\prime2}}{M_\chi^\dagger M_\chi}\))_{\rm LR}
\frac1{M_\chi}\right\}_{\rm hh}\bm m_u\bm K\right]_{11},
 \label{cha}
\ee
where
the subscript LR means the left-right block of the scalar-quark mass
matrix (Eq.\ (\ref{scalarmass}))
and the subscript 11 denotes the (1,1) component
in the generation indices.
The subscript hh denotes the (2,2) component
in the wino-Higgsino mass matrix  $M_\chi$ which is given by
\be
 -\barr{cc}
   \overline{{\widetilde W}_R^-}&\overline{{\tilde h}_{uR}^-}
  \earr
                                          M_\chi
                                                \barr c
                                                 \widetilde W_L^-\\
                                                 \tilde h_{dL}^-
                                                \earr
                                                        +{\rm H.c.},
\ee
\be
 M_\chi\equiv\barr{cc}M_2                    &\sqrt 2M_W\cos\beta\\
                   \sqrt 2M_W\sin\beta       &-\mu            \earr.
\ee
The function $F(x)$ is given by
\be F(x)\equiv \frac1{6 (1-x)^3} (5-12 x+7 x^2+2 x (2-3 x) \log x).\ee

To evaluate the expression Eq.\ (\ref{cha}), we need to diagonalize
the scalar-quark mass matrix $M_{\tilde u}^{\prime2}$ and
the wino-Higgsino one $M_\chi$.
The former is explicitly diagonalized by a $6\times 6$
unitary matrix $U_{\tilde u}$,
\be
U_{\tilde u} M_{\tilde u}^{\prime2} U_{\tilde u}^{\dagger}
={\hat M}_{\tilde u}^2.
\label{scalardiag}
\ee
The ${\hat M}_{\tilde u}^2$ is a diagonal matrix whose elements are
all positive.
On the other hand, the latter can be analytically diagonalized by
two orthogonal matrices $U_R$ and $U_L$ as
\be
U_R M_\chi U_L^T=
\barr{cc}
m_{\chi^1} & 0\\
0 & m_{\chi^2}
\earr,
\label{chadiag}
\ee
where the mass of light (heavy) chargino is denoted as $m_{\chi^1}$
($m_{\chi^2}$)
\ba
m_{\chi^1},\;m_{\chi^2}&=&\frac 12
\left|\sqrt{(M_2-\mu)^2+2M_W^2(1-\sin 2\beta)}\right. \nonumber \\
&&\left.
\;\;\;\;\mp\sqrt{(M_2+\mu)^2+2M_W^2(1+\sin 2\beta)}\right|.
\ea
Using those matrices,
Eq.\ (\ref{cha}) can be rewritten as follows,
\be
d_d=(U_L^T)_{2a}D(m_{\chi^a})(U_R)_{a2},
\label{chacha}
\ee
where $a=1,\;2$ denotes the mass eigenstates as in
Eq.\ (\ref{chadiag}).
The function $D(m_{\chi^a})$ is given by
\be
D(m_{\chi^a})=
\frac e{16\pi^2}(\tan\beta+\cot\beta)
 \frac2{v^2}\frac{m_d}{m_{\chi^a}}\,
{\rm Im}\!\left[\bm K^\dagger
\left\{U_{\tilde u}^\dagger F\!\((
\frac{\hat{M}_{\tilde u}^2}{m_{\chi^a}^2}\))
U_{\tilde u}\right\}_{\rm LR}\bm m_u\bm K\right]_{11}
\ee
Since we can express the function $F(x)$ and the
orthogonal matrices $U_R$ and $U_L$ in terms of eigenvalues
$m_{\chi^a}$ and $\hat{M}_{\tilde u}$, we obtain
the EDM of Eq.\ (\ref{chacha}) as
\ba
d_d&\!\!\!=&\!\!\!\pm \frac 12
\left[\frac{D(m_{\chi^1})+D(m_{\chi^2})}{m_{\chi^1}+m_{\chi^2}}
\sqrt{(m_{\chi^1}+m_{\chi^2})^2-2M_W^2(1\mp\sin 2\beta)}
\right. \\
&\!\!\!&\!\!\!
\left.
+
{\rm sgn}(M_2- |\mu|)
\frac{D(m_{\chi^1})-D(m_{\chi^2})}{m_{\chi^2}-m_{\chi^1}}
\sqrt{(m_{\chi^1}-m_{\chi^2})^2-2M_W^2(1\pm\sin 2\beta)}
\right],\nonumber
\ea
where the upper (lower) sign corresponds to the positive (negative)
sign of $\det M_{\chi}$.

Since the off-diagonal elements of the $\bm{A}$ matrix are much
smaller than the diagonal ones in magnitude,
let us first examine the effect other than the off-diagonal elements
of the $\bm{A}$ matrix.
Namely we tentatively assume that $\bm{A}$ is just a number $A$
without the off-diagonal elements.
Then the scalar-quark mass matrices have off-diagonal elements in
generation indices only in the LL blocks.
In this case, many diagrams become real and do not contribute to
EDM, since
the phases coming from the Cabibbo-Kobayashi-Maskawa
matrix are canceled due to the hermiticity of mass matrices.
Without the off-diagonal elements, we find the contribution of this
chargino diagram to give the EDM
of the order of $10^{-31}\sim 10^{-33}\: e\cdot$cm,
assuming SUSY mass parameters are of the order of 100 GeV.

Next let us examine the effect of the off-diagonal elements of
the matrix $\bm{A}$.
The scalar-quark mass matrices at low energies
after the renormalization group running can be
represented as Eq.\ (\ref{scalarmass}) in the super KM basis.
The off-diagonal elements of the $\bm{A}$ matrices
are small in magnitude but have
complex phases of $O(1)$ as in
Eqs.\ (\ref{eqn:valuealu}) and (\ref{eqn:valueamu}).
{}Furthermore they provide new sources of
flavor changing neutral current.
These flavor changing currents spoil the cancellation of the KM
phases in the one-loop diagrams for the EDM.
Therefore the structure of the LL blocks of scalar-quark mass matrices
are not important in this case.

The presence of the off-diagonal elements of $\bm{A}$ generally
helps to give a larger contribution to EDM.
As shown in Fig.~\ref{chrg},
the contribution of the chargino
diagram to the EDM in fact increases and becomes
of the order of $10^{-27} \sim 10^{-29}\ e \cdot \rm{cm}$.
In Fig.~\ref{chrg}, we have chosen $\tan \beta=10$ and $m_{\chi^1}=50$
GeV.
One should note that EDM is approximately proportional to $\tan \beta$
for $\tan \beta >5$.
In order to see the allowed parameter region, we have plotted in
{}Fig.~\ref{region} the region of positive squared masses for
scalar-quarks.
The allowed parameter regions differ depending on $M_2 > |\mu|$
(Fig.~4a) or $M_2 < |\mu|$ (Fig.~4b).

Since d-type-scalar-quarks instead of u-type-scalar-quarks are
involved in the loop diagram, we find that the EDM
$d_u$ of u-quarks is smaller than that of d-quarks and is
of the order of $10^{-33}\sim 10^{-35}\: e\cdot$cm.

Next let us examine the gluino contributions to the neutron EDM
which is shown in Fig.~\ref{glud},
\be
 d_d=\frac e{16\pi^2}g_3^{\:2}\frac43\frac1{M_3}
{\rm Im}\!\((G\!\((\frac{M_{\tilde d}^{\prime2}}{M_3^{\:2}}\))_{\rm
    RL}\))_{11},
\ee
where $g_3$ and $M_3$ are the strong interaction coupling constant
and the mass of the gluino respectively and
\be
G(x)\equiv \frac1{3(1-x)^3} (1-x^2+2 x \log x).
\ee

Diagonalizing explicitly the d-type-scalar-quark mass matrix, we find
that the EDM $d_d$ of the down quark have contributions from this
gluino diagram of the order of $10^{-34}\sim 10^{-35}\: e\cdot$cm.
The complex phases
in off-diagonal elements of $A$ terms
are the major source of this contribution.

In the $SU(6)$ quark model, the neutron EDM is given in terms of the
single quark EDM as
\be
d_n={4 \over 3}d_d-{1 \over 3}d_u .
\ee
{}From the above results, we find that
the neutron EDM from the single quark EDM is of the order of
$10^{-27}\sim 10^{-29}\: e\cdot$cm for $\tan \beta=10$.
The chargino
diagram of Fig.~\ref{chrd}
is the
dominant contribution,
since $\bm A$ terms have off-diagonal elements.
Although gluino diagrams can also contribute to EDM,
its contribution is smaller than that of
chargino,
since complex phases are almost canceled if one takes
the (1,1) component in generation indices.

The contribution of chargino diagram in Fig.~\ref{chrd} was
also examined
in ref.\ \cite{BV} with somewhat different values of parameters
and a result of
the order of $10^{-30}\: e\cdot$cm for the EDM of neutron was
reported.
We have taken account of the evolution of gaugino masses in our
renormalization group analysis.
This may be a reason for the fact that we have obtained a larger value
for the neutron EDM in comparison to ref.\ \cite{BV}.


\newsection{Effects of the transition electric dipole moment}

It has been shown that the neutron EDM is also induced in the
nonsupersymmetric minimal standard model through the
transition electric dipole moment (TEDM)
of quarks within the baryon as illustrated in Fig.~\ref{tedmd}
\cite{Morel,GAV,GAVpenguin}.
In this section, we examine the TEDM, {$d \rightarrow s+\gamma$},
in the MSSM and calculate the neutron EDM based on their method.
The matrix element of the electromagnetic current $j^{\mu}$
between single particle states of different Dirac particles can be
written as follows,
\ba
 \langle\bm p_f,s_f|j^\mu(0)|\bm p_i,s_i\rangle\nonumber
  &=&\bar u_f (\bm p_f,s_f)
     \left[i\sigma^{\mu\nu}q_\nu\frac{F_2(q^2)}{m_f+m_i}
           +\gamma_5\sigma^{\mu\nu}q_\nu\frac{F_3(q^2)}{m_f+m_i}
        \right.\nonumber\\
  & &\qquad\qquad\quad+\{q^2\gamma^\mu-(m_f-m_i)q^\mu\}F_4(q^2)\\
  & &\qquad\qquad\quad+\left.\((\frac{q^2}{m_f+m_i}\gamma^\mu-q^\mu\))
         \gamma_5F_A(q^2)\right]u_i (\bm p_i,s_i),\nonumber
\ea
where $m_i$ and $m_f$ are initial and final Dirac fermion masses
respectively.
The
dominant contribution to EDM comes from $F_3$ term as that of single
quark case
when the mass difference $|m_f-m_i|$ is small enough compared to the
other relevant masses.
In our particular case, the operator relevant to the TEDM can be
written as
\be
\kappa^{sd} \bar{s}(p_s) i \sigma_{\mu \nu}
                (p_s-p_d)^{\nu} \gamma_5 d(p_d),
\ee
where
\be i\kappa^{sd}\approx\frac{F_3(0)}{m_s+m_d}.\ee
Im $\kappa^{sd}$ is just the TEDM.

The TEDM in the nonsupersymmetric minimal
standard model has already been calculated
under the assumption that $m_t \ll M_W$ \cite{Morel,GAV}.
Recently it is more and more certain experimentally
that the top quark mass is very large \cite{CDF}.
Therefore we
must redo the calculation in the nonsupersymmetric case
by taking into account of the large top quark mass.
Assuming $m_d$ and $m_s$ to be small in comparison with the masses
of the internal lines of the loop diagram,
we have the following contribution to the
TEDM of $d \rightarrow s+\gamma$,
\be
\kappa^{sd}=\frac{G_{F}}{\sqrt2} \frac e{(4 \pi)^2}K_{ts}^*K_{td}
             (m_{d}-m_{s})\: f\!\left(\frac{m_t^2}{M_W^{\ 2}}\right),
\ee
where \cite{INAMILIM}
\be
f(x)\equiv\frac x{6(x-1)^4} (-8x^3+3x^2+12x-7+6x(3x-2) \log x).
\ee
{}From the above equation, we obtain that
the standard model gives $2\cdot10^{-26}\: e\cdot$cm for the TEDM
using the Cabibbo-Kobayashi-Maskawa matrix (\ref{ckmmatrix}) --
(\ref{delta}).
Since the hadronic effects to convert the TEDM
to the neutron EDM give a factor of $10^{-7}$ \cite{GAV},
the contribution from the TEDM in the standard model
to the neutron EDM becomes of the order of $10^{-33}
\: e\cdot$cm.

Next let us consider the TEDM of quarks in the MSSM.
Similarly to the quark EDM,
the chargino and gluino diagrams
are most important.
Since the bound state effects should be the same for supersymmetric
and nonsupersymmetric models,
we shall calculate these diagrams for the quark TEDM
and compare them with those of the standard model.

In order to obtain the TEDM, we have only to replace $d_{L}$
by $s_{L}$ in the analysis in the previous section.
Similarly to the diagram in Fig.~\ref{chrd},
for instance,
the chargino contribution
is obtained
from Eq.~(\ref{cha}) as
\be
 {\rm Im}\: \kappa^{sd}=\frac e{16\pi^2}(\tan\beta+\cot\beta)
\frac2{v^2}m_d\,{\rm Im}\!
\left[\bm K^\dagger\left\{F\!\((
\frac{M_{\tilde u}^{\prime2}}{M_\chi^\dagger M_\chi}\))_{\rm LR}
\frac1{M_\chi}\right\}_{\rm hh}\bm m_u\bm K\right]_{12}.
 \label{chatr}
\ee
Let us note that
the element we are interested in is not (2,1) but (1,2)
for the LR block of the scalar-quark mass matrix,
since the Higgsino coupling changes chirality such as
{$\tilde{u}^i_R \rightarrow s_L$.
By the same token, the gluino diagram for the quark TEDM corresponding
to  the diagram in Fig.~\ref{glud} is obtained as
\be
 {\rm Im}\:\kappa^{sd}=\frac e{16\pi^2}g_3^{\:2}\frac43\frac1{M_3}
{\rm Im}\!\((G\!\((\frac{M_{\tilde d}^{\prime2}}{M_3^{\:2}}\))
_{\rm RL}\))_{12}.
 \label{glutr}
\ee
In this case, we take the (1,2) element for the RL block because
gauge couplings do not change chirality.

We obtain the TEDM of the order of $10^{-25}\sim 10^{-27}\: e\cdot$cm
for the chargino contribution as shown in Fig.~\ref{tedmresult}
and $10^{-28}\sim 10^{-30}\: e\cdot$cm for the gluino one.
Thus we find that the TEDM in the MSSM is of the same order of
magnitude as that in the nonsupersymmetric standard model.
By combining our results with the hadronic matrix elements
estimated already in the nonsupersymmetric case,
we find that
the resulting neutron EDM becomes of the order of
$10^{-32}\sim 10^{-34}\: e\cdot$cm.

Let us also consider the effects of other diagrams in the
supersymmetric model.
Besides the diagram which can be considered as the quark TEDM,
there are diagrams where two or more quarks within the neutron
exchange the SUSY particles.
The R-parity conservation in the MSSM dictates that such diagrams
must be box diagrams where all of the internal particles are SUSY
ones.  So the intermediate states at the hadronic level are of the
order of mass scale of the SUSY particles which are much heavier than
those in the diagrams that we have considered.
Therefore the resulting neutron EDM is expected to be highly
suppressed.

Apart from the TEDM that we have considered in Fig.~\ref{tedmd},
another proposal for an effect involving many quarks
was made in ref.\ \cite{GAVpenguin}.
They considered the so-called penguin diagrams for the TEDM of quarks.
They found that the conversion factor from TEDM to the neutron EDM is
$10^{-4}$ instead of $10^{-7}$ due to a large QCD enhancement.
Therefore this diagram gives the neutron EDM of the order of
$10^{-30}\:e\cdot$cm which is the largest
contribution in the nonsupersymmetric standard model.
Since the enhancement due to the QCD corrections
is the same order of magnitude in the MSSM
as in the nonsupersymmetric standard model,
we obtain that the neutron EDM is of the order of
$10^{-29}\sim 10^{-31}\: e\cdot$cm.
Moreover we find from Fig.~\ref{tedmresult}
that the TEDM is of the same order of magnitude even
for small values of $A$ such as $A<1$ TeV, whereas the single quark
EDM becomes very small as shown in Fig.~\ref{chrg}.
Therefore the penguin diagram with TEDM is more important than a
single quark EDM for smaller values of $A$
($A\stackrel{<}{\sim} 1$ TeV).

\vspace{5mm}


One of the authors (N.S.) thanks to Y. Okada, T. Goto and J. Hisano
for a useful discussion on supersymmetric models and flavor-changing
neutral currents.
This work is supported in part by Grant-in-Aid for
Scientific Research (T.I. and Y.M.) and (No.05640334) (N.S.), and
Grant-in-Aid for Scientific Research for Priority Areas
(No.06221222) (N.S.) from the Ministry of Education, Science
and Culture.

\renewcommand{\thesection}{A}
\section{Renormalization Group Equations}\label{adx:rge}
\setcounter{equation}{0}
\renewcommand{\theequation}{A.\arabic{equation}}

\newcommand\alp{\tilde{\bm\alpha}}
\newcommand\alpt{\tilde\alpha}
\newcommand\Y{\widetilde{\bm Y}}

  We define the scaling variable $t$ using the GUT scale $M_{GUT}$ and
  the relevant momentum $Q$ as
\be t\equiv\log \frac{M_{GUT}^{\ 2}}{Q^2}. \ee
A tilde over couplings denotes a division by a factor $4\pi$, namely
$\alpt\equiv\alpha/(4\pi)$.  The renormalization group equations for
the gauge couplings and gaugino masses $M_i$ are given as follows
\cite{BBMR}:
\ba \dot{\alpt}_i&=&-b_i\alpt_i^{\ 2},\\
    \dot{M}_i    &=&-b_i\alpt_iM_i,
\ea
where
\ba b_1=\frac{33}5,&b_2=1,&b_3=-3, \ea
and we denote derivative by $t$ with a dot.
The solutions are given as follows:
\be
 \alpt_i(t)=\frac{\alpt_i(t_0)}{1+b_i(t-t_0)\alpt_i(t_0)}
                  =\frac{\alpt_G}{1+b_it\alpt_G},
\ee
\be
 M_i(t)=\frac{M_i(t_0)}{1+b_i(t-t_0)\alpt_i(t_0)}
       =\frac{\alpt_i(t)}{\alpt_i(t_0)}M_i(t_0)
       =\frac M{\alpt_G}\alpt_i(t),
\ee
where we assume the gauge coupling unification
and the universal gaugino mass
at the GUT scale:
\be \alpha_1(0)=\alpha_2(0)=\alpha_3(0)=\alpha_G, \ee
\be M_1(0)=M_2(0)=M_3(0)=M. \ee

The renormalization group equations for Yukawa couplings are written
as follows \cite{BBMR}:
\ba
 2\dot{\Y}_u&=&
  \((\frac{13}{15}\alpt_1+3\alpt_2+\frac{16}3\alpt_3\))\Y_u
  -\left[3\Y_u\Y_u^\dagger\Y_u+3\Tr\((\Y_u\Y_u^\dagger\))\Y_u\right]
  \nonumber\\
            & & -\Y_u\Y_d^\dagger\Y_d,\\
 2\dot{\Y}_d&=&
  \((\frac{7}{15}\alpt_1+3\alpt_2+\frac{16}3\alpt_3\))\Y_d
  -\left[3\Y_d\Y_d^\dagger\Y_d+3\Tr\((\Y_d\Y_d^\dagger\))\Y_d\right]
  \nonumber\\
            & &-\Y_d\Y_u^\dagger\Y_u-\Tr\((\Y_e\Y_e^\dagger\))\Y_d,\\
 2\dot{\Y}_e&=&
  \((\frac{9}{5}\alpt_1+3\alpt_2\))\Y_e
  -\left[3\Y_e\Y_e^\dagger\Y_e+\Tr\((\Y_e\Y_e^\dagger\))\Y_e\right]
  \nonumber\\
            & &-3\Tr\((\Y_d\Y_d^\dagger\))\Y_e.
\ea

We now define hermitian matrices $\alp_f$ ($f=u,d,e$),
\be \alp_f\equiv\Y_f^\dagger\Y_f. \ee
The renormalization group equations for $\alp_f$ are given as follows:
\ba
 2\dot{\alp}_u&=&
  2\((\frac{13}{15}\alpt_1+3\alpt_2+\frac{16}3\alpt_3\))\alp_u
  -2\left[3\alp_u^{\ 2}+3\Tr\((\alp_u\))\alp_u\right]\nonumber\\
              & &-\alp_d\alp_u-\alp_u\alp_d,\\
 2\dot{\alp}_d&=&
  2\((\frac{7}{15}\alpt_1+3\alpt_2+\frac{16}3\alpt_3\))\alp_d
  -2\left[3\alp_d^{\ 2}+3\Tr\((\alp_d\))\alp_d\right]
  -2\Tr\((\alp_e\))\alp_d\nonumber\\
              & &-\alp_u\alp_d-\alp_d\alp_u,\\
  \dot{\alp}_e&=&
  \((\frac{9}{5}\alpt_1+3\alpt_2\))\alp_e
  -\left[3\alp_e^{\ 2}+\Tr\((\alp_e\))\alp_e\right]
                                   -3\Tr\((\alp_d\))\alp_e.
\ea

The renormalization group equations for $\bm A$'s can be written as
follows \cite{BBMR,BV}:
\ba
 2\dot{\bm A}_u&=&
  -2\((\frac{13}{15}\alpt_1M_1+3\alpt_2M_2
                            +\frac{16}3\alpt_3M_3\))\one
  -2\Tr\((3\bm A_u\alp_u\))\one\nonumber\\
               & &-5\alp_u\bm A_u-\bm A_u\alp_u-\alp_d\bm A_u+\bm
                                       A_u\alp_d-2\bm A_d\alp_d,\\
 2\dot{\bm A}_d&=&
  -2\((\frac{7}{15}\alpt_1M_1+3\alpt_2M_2+\frac{16}3\alpt_3M_3\))\one
  -2\Tr\((\bm A_e\alp_e+3\bm A_d\alp_d\))\one\nonumber\\
               & &-5\alp_d\bm A_d-\bm A_d\alp_d-\alp_u\bm A_d+\bm
                                       A_d\alp_u-2\bm A_u\alp_u,\\
  \dot{\bm A}_e&=&
  -2\((\frac{9}{5}\alpt_1M_1+3\alpt_2M_2\))\one
  -2\Tr\((\bm A_e\alp_e+3\bm A_d\alp_d\))\one\nonumber\\
               & &-5\alp_e\bm A_e-\bm A_e\alp_e.
\ea

Next we consider the renormalization group equations in the
super KM basis, i.e., we rotate Yukawa couplings $\Y_f$ by the
same amount as $\bm M_f$ in Eqs.\
(\ref{eqn:rotatemu}) and (\ref{eqn:rotatemd}),  and $\bm A_f$ and
$\alp_f$ by the same amount as
$\bm A_f$ in Eq.\ (\ref{eqn:rotatea}).  In this basis the
renormalization group equations are given as follows:
\ba
 2\dot{\alp}_u&=&
  2\((\frac{13}{15}\alpt_1+3\alpt_2+\frac{16}3\alpt_3\))\alp_u
  -2\left[3\alp_u^{\ 2}+3\Tr\((\alp_u\))\alp_u\right]\nonumber\\
              & &-\bm K\alp_d\bm K^\dagger\alp_u
                       -\alp_u\bm K\alp_d\bm K^\dagger,
                                          \label{eqn:rgealphau}\\
 2\dot{\alp}_d&=&
  2\((\frac{7}{15}\alpt_1+3\alpt_2+\frac{16}3\alpt_3\))\alp_d
  -2\left[3\alp_d^{\ 2}+3\Tr\((\alp_d\))\alp_d\right]
  -2\Tr\((\alp_e\))\alp_d\nonumber\\
              & &-\bm K^\dagger\alp_u\bm K\alp_d
                -\alp_d\bm K^\dagger\alp_u\bm K,\\
  \dot{\alp}_e&=&
  \((\frac{9}{5}\alpt_1+3\alpt_2\))\alp_e
  -\left[3\alp_e^{\
    2}+\Tr\((\alp_e\))\alp_e\right]-3\Tr\((\alp_d\))\alp_e,
                                       \label{eqn:rgealphae}
\ea

\ba
2\dot{\bm A}_u&=&
  -2\((\frac{13}{15}\alpt_1M_1+3\alpt_2M_2+\frac{16}3\alpt_3M_3\))\one
  -2\Tr\((3\bm A_u\alp_u\))\one\nonumber\\
&&\!\!\!\!\!\!-5\alp_u\bm A_u-\bm A_u\alp_u
 -\bm K\alp_d\bm K^\dagger\bm A_u+\bm A_u\bm K\alp_d\bm K^\dagger
    -2\bm K\bm  A_d\alp_d\bm K^\dagger,
                                                 \label{eqn:rgeau}\\
 2\dot{\bm A}_d&=&
  -2\((\frac{7}{15}\alpt_1M_1+3\alpt_2M_2+\frac{16}3\alpt_3M_3\))\one
  -2\Tr\((\bm A_e\alp_e+3\bm A_d\alp_d\))\one\nonumber\\
&&\!\!\!\!\!\!-5\alp_d\bm A_d-\bm A_d\alp_d
 -\bm K^\dagger\alp_u\bm K\bm A_d+\bm A_d\bm K^\dagger\alp_u\bm K
     -2\bm K^\dagger\bm A_u\alp_u\bm K,\\
 \dot{\bm A_e}&=&
  -2\((\frac{9}{5}\alpt_1M_1+3\alpt_2M_2\))\one
  -2\Tr\((\bm A_e\alp_e+3\bm A_d\alp_d\))\one\nonumber\\
               & &-5\alp_e\bm A_e-\bm A_e\alp_e.\label{eqn:rgeae}
\ea

In Eqs.\ (\ref{eqn:rgealphau}) -- (\ref{eqn:rgealphae}), we give
the initial conditions for $\alp_f$ ($f=u,d,e$) at the scale of
$Z^0$ boson mass $M_Z$
which are obtained from the quark and lepton masses, i.e.,
\ba
 \alp_u&=&\((\frac{\sqrt2}{4\pi v\sin\beta}\))^2\bm m_u^{\ 2},\\
 \alp_d&=&\((\frac{\sqrt2}{4\pi v\cos\beta}\))^2\bm m_d^{\ 2},\\
 \alp_e&=&\((\frac{\sqrt2}{4\pi v\cos\beta}\))^2\bm m_e^{\ 2}
\ea
at the scale $M_Z$ and we obtain solutions for $\alp_f$.  The
$\alp_f$ are diagonal at the initial condition and then $\alp_e$ is
diagonal at all the scale but $\alp_u$ and $\alp_d$ are not diagonal
at the higher scale.

In Eqs.\ (\ref{eqn:rgeau}) -- (\ref{eqn:rgeae}), we give the initial
conditions for $\bm A_f$ ($f=u,d,e$) at the GUT scale $M_{GUT}$ which
are universal, i.e.,
\be \bm A_u(0)=\bm A_d(0)=\bm A_e(0)=A\one. \label{eqn:ainitial}\ee
The $\bm A_f$ are diagonal at the initial condition.  Then $\bm A_e$
is diagonal at all the scale, but $\bm A_u$ and $\bm A_d$ are not
diagonal at the lower scale.  It is convenient to separate $\bm A_f$
into two parts.
\be \bm A_f=\bm A_{Lf}+\bm A_{Mf}, \ee
where $f=u,d,e$.
The same renormalization group equations
 (\ref{eqn:rgeau}) -- (\ref{eqn:rgeae}) as $\bm A_f$
are valid for $\bm A_{Mf}$ which are
proportional to the universal gaugino mass $M$.
Their initial conditions at the GUT scale are
\be \bm A_{Mu}(0)=\bm A_{Md}(0)=\bm A_{Me}(0)= 0. \ee
By deleting gaugino masses from
Eqs.\ (\ref{eqn:rgeau}) -- (\ref{eqn:rgeae}), we obtain
linear equations for
$\bm A_{Lf}$:
\ba
 2\dot{\bm A}_{Lu}&=&
  -2\Tr\((3\bm A_{Lu}\alp_u\))\one-5\alp_u\bm A_{Lu}
                         -\bm A_{Lu}\alp_u\nonumber\\
   &&-\bm K\alp_d\bm K^\dagger\bm A_{Lu}
         +\bm A_{Lu}\bm K\alp_d\bm K^\dagger-2\bm K\bm
   A_{Ld}\alp_d\bm K^\dagger,\label{eqn:rgealu}\\
 2\dot{\bm A}_{Ld}&=&
  -2\Tr\((\bm A_{Le}\alp_e+3\bm A_{Ld}\alp_d\))\one
   -5\alp_d\bm A_{Ld}-\bm A_{Ld}\alp_d\nonumber\\
                  & &-\bm K^\dagger\alp_u\bm K\bm A_{Ld}
                         +\bm A_{Ld}\bm K^\dagger\alp_u\bm K
                     -2\bm K^\dagger\bm A_{Lu}\alp_u\bm K,\\
  \dot{\bm A}_{Le}&=&
  -2\Tr\((\bm A_{Le}\alp_e+3\bm A_{Ld}\alp_d\))\one
   -5\alp_e\bm A_{Le}-\bm A_{Le}\alp_e.\label{eqn:rgeale}
\ea
The initial conditions are the same as $\bm A_f$ and
are proportional to $A$, i.e.,
\be \bm A_{Lu}(0)=\bm A_{Ld}(0)=\bm A_{Le}(0)=A\one. \ee
We consider $m_t=174$ GeV \cite{CDF}.
The Cabibbo-Kobayashi-Maskawa matrix is parametrized as \cite{PARTPR}
\be
\bm K=\barr{ccc}
 c_2c_3&c_2s_3&s_2e^{-i\delta}\\
 -c_1s_3-s_1s_2c_3e^{i\delta}&c_1c_3-s_1s_2s_3e^{i\delta}&s_1c_2\\
 s_1s_3-c_1s_2c_3e^{i\delta}&-s_1c_3-c_1s_2s_3e^{i\delta}&c_1c_2
  \earr,
\label{ckmmatrix}
\ee
where
\be c_i=\cos\theta_i,\qquad s_i=\sin\theta_i\qquad (i=1,2,3).\ee
We use the following angles from phenomenological analyses
\cite{PARTPR}
\be
 \theta_1=0.043, \quad
 \theta_2=0.005, \quad
 \theta_3=0.221, \quad
 \delta  =0.86.
\label{delta}
\ee
{}For $\tan\beta=10$ and $A=1$ GeV, we obtain $A_{Lf}$
at the scale $M_Z$ by solving the renormalization group equations:
\be
 \bm A_{Lu}=\barr{ccc}
  0.60 &       (-5.4+i1.8)\cdot10^{-6}&(-3.6+i4.2)\cdot10^{-5}\\
  (-5.4-i1.8)\cdot10^{-6}&0.60        &-4.8\cdot10^{-4}       \\
  (-3.6-i4.2)\cdot10^{-5}&-4.8\cdot 10^{-4}       &0.19
  \earr,\label{eqn:valuealu}
\ee
\be
 \bm A_{Ld}=\barr{ccc}
  0.96 &       (3.7+i2.2)\cdot10^{-5}&(-8.3-i4.9)\cdot10^{-4}\\
  (3.7-i2.2)\cdot10^{-5} &0.96       &(5.7-i0.1)\cdot 10^{-3}\\
  (-8.3+i4.9)\cdot10^{-4}&(5.7+i0.1)\cdot10^{-3}&0.79
  \earr,
\ee
\be
 \bm A_{Le}=\barr{ccc}0.96&0   &0   \\
                      0   &0.96&0   \\
                      0   &0   &0.95\earr.
\ee
{}For $\tan\beta=10$ and $M=1$ GeV, we obtain $A_{Mf}$
at the scale $M_Z$:
\be
 \bm A_{Mu}=\barr{ccc}
  -2.9               &(1.1-i0.4)\cdot10^{-5}&(7.6-i8.9)\cdot10^{-5}\\
  (1.1+i0.4)\cdot10^{-5}&-2.9               &0.0010                \\
  (7.6+i8.9)\cdot10^{-5}&0.0010             &-2.1           \earr,
\label{eqn:valueamu}
\ee
\be
 \bm A_{Md}=\barr{ccc}
  -3.7              &(-8.0-i4.7)\cdot10^{-5}&(1.8-i1.1)\cdot10^{-3}\\
  (-8.0+i4.7)\cdot10^{-5}&-3.7              &-0.012                \\
  (1.8+i1.1)\cdot10^{-3} &-0.012            &-3.3
  \earr,
\ee
\be
 \bm A_{Me}=\barr{ccc}-0.62&0   &0    \\
                      0   &-0.62&0    \\
                      0   &0    &-0.61\earr.\label{eqn:valueame}
\ee

\vspace{5mm}

\vspace{5mm}
\newpage

\newcommand{\NP}[1]{{\it Nucl.\ Phys.}\/\ {\bf #1}}
\newcommand{\PL}[1]{{\it Phys.\ Lett.}\/\ {\bf #1}}
\newcommand{\CMP}[1]{{\it Commun.\ Math.\ Phys.}\/\ {\bf #1}}
\newcommand{\MPL}[1]{{\it Mod.\ Phys.\ Lett.}\/\ {\bf #1}}
\newcommand{\IJMP}[1]{{\it Int.\ J. Mod.\ Phys.}\/\ {\bf #1}}
\newcommand{\PR}[1]{{\it Phys.\ Rev.}\/\ {\bf #1}}
\newcommand{\PRL}[1]{{\it Phys.\ Rev.\ Lett.}\/\ {\bf #1}}
\newcommand{\PTP}[1]{{\it Prog.\ Theor.\ Phys.}\/\ {\bf #1}}
\newcommand{\PTPS}[1]{{\it Prog.\ Theor.\ Phys.\ Suppl.}\/\ {\bf #1}}
\newcommand{\AP}[1]{{\it Ann.\ Phys.}\/\ {\bf #1}}

\clearpage

\section*{Figure captions}
\begin{itemize}
\item[Fig.\ 1]
The chargino contribution to the quark EDM involving Higgsino
couplings.
The arrow on each line stands for the chirality
of the particle and the cross for mass insertion.
\item[Fig.\ 2]
The gluino contribution to the quark EDM.
\item[Fig.\ 3]
{}For the case $\det M_\chi>0$, EDM of the down quark is
   plotted as a function of the mass $m_{\chi^2}$ of the
   heavier chargino for various
   values of $A$ in the case (a) $M_2+\mu>0$ and (b) $M_2+\mu<0$.
   We have chosen parameters: universal
   scalar-mass $m=100$ GeV, $\tan \beta =10$, and the mass of the
light
   chargino $m_{\chi^1}=50$ GeV.
\item[Fig.\ 4]
The region of parameters for the squared mass of scalar-quarks
   to be positive, for (a) $M_2+\mu>0$ and (b) $M_2-|\mu|<0$.
   The boundary of vanishing mass of u-type-scalar-quark is
   represented by solid lines, and that of d-type by dashed lines.
   In the case (a), allowed region for $m=100$ GeV, $\mu<0$, and
   $\tan \beta=10$ is denoted by shaded area. For $\tan \beta=100$,
   the allowed region is the right of dotted line bounded by two solid
   lines.
   In the case (b), allowed region for $\mu>0$
   is shown in the upper half plane and that for $\mu<0$ in the lower
   half. Allowed region for $m=100$ GeV and $\tan \beta=10$ is denoted
   by shaded area.
\item[Fig.\ 5]
Contribution to the neutron EDM through the TEDM.
The diagrams like Fig.~\ref{chrd} or Fig.~\ref{glud} is inserted
in the blob.
\item[Fig.\ 6]
{}For the case $\det M_\chi>0$, TEDM ($d\rightarrow s\gamma$) is
   plotted as a function of the mass $m_{\chi^2}$ of the
   heavier chargino for various
   values of $A$ in the case (a) $M_2+\mu>0$ and (b) $M_2+\mu<0$.
   We have chosen parameters: universal
   scalar mass $m=100$ GeV, $\tan \beta =10$, and the mass of the
light
   chargino $m_{\chi^1}=50$ GeV.
\end{itemize}

\begin{figure}
 \leavevmode
 \epsfysize=5cm
 \centerline{\epsfbox{chrg.eps}}

\vspace{3mm}

 \epsfysize=5cm
 \centerline{\epsfbox{chrg2.eps}}
 \caption{
The chargino contribution to the quark EDM involving Higgsino
couplings.
The arrow on each line stands for the chirality
of the particle and the cross for mass insertion.
}
 \label{chrd}

\vspace{3mm}

 \leavevmode
 \epsfysize=5cm
 \centerline{\epsfbox{glu.eps}}
 \caption{The gluino contribution to the quark EDM.}
 \label{glud}

\end{figure}

\begin{figure}
 \leavevmode
 \epsfysize=8cm
 \centerline{\epsfbox{fig3a.eps}}

 \leavevmode
 \epsfysize=8cm
 \centerline{\epsfbox{fig3b.eps}}
 \caption{
{}For the case $\det M_\chi>0$, EDM of the down quark is
   plotted as a function of the mass $m_{\chi^2}$ of the
   heavier chargino for various
   values of $A$ in the case (a) $M_2+\mu>0$ and (b) $M_2+\mu<0$.
   We have chosen parameters: universal
   scalar mass $m=100$ GeV, $\tan \beta =10$, and the mass of the
   light chargino $m_{\chi^1}=50$ GeV.
}
 \label{chrg}
\end{figure}

\begin{figure}
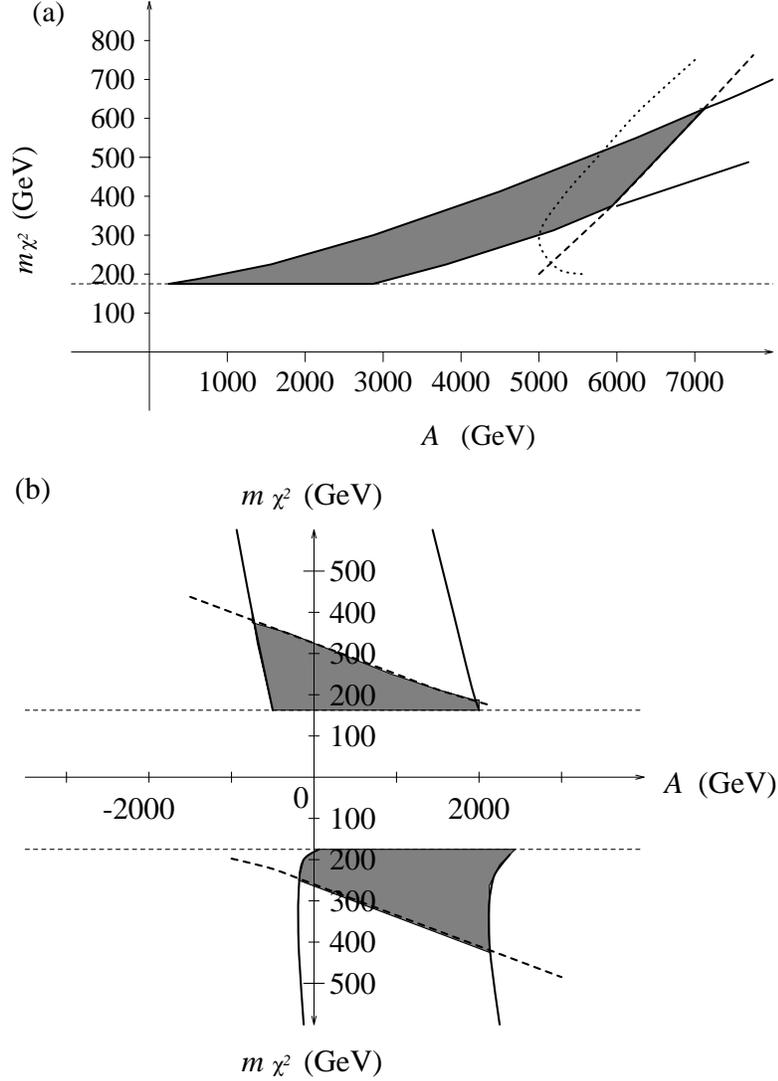

\leavevmode
 \epsfysize=6cm
 \centerline{\epsfbox{fig4a.eps}}

\vspace{3mm}

\leavevmode
 \epsfysize=8cm
 \centerline{\epsfbox{fig4b.eps}}
 \caption{The region of parameters for the squared mass of
 scalar-quarks
   to be positive, for (a) $M_2+\mu>0$ and (b) $M_2-|\mu|<0$.
   The boundary of vanishing mass of u-type-scalar-quark is
   represented by solid lines, and that of d-type by dashed lines.
   In the case (a), allowed region for $m=100$ GeV, $\mu<0$, and
   $\tan \beta=10$ is denoted by shaded area. For $\tan \beta=100$,
   the allowed region is the right of dotted line bounded by two solid
   lines.
   In the case (b), allowed region for $\mu>0$
   is shown in the upper half plane and that for $\mu<0$ in the lower
   half. Allowed region for $m=100$ GeV and $\tan \beta=10$ is denoted
   by shaded area.}
 \label{region}
\end{figure}

\begin{figure}
 \leavevmode
 \epsfysize=5cm
 \centerline{\epsfbox{tedm.eps}}
 \caption{Contribution to the neutron EDM from the TEDM.
          The diagrams like Fig.\ 1 or Fig.\ 2 are inserted
          in the blob.}
 \label{tedmd}

 \vspace{3mm}

 \leavevmode
 \epsfysize=6cm
 \centerline{\epsfbox{fig6a.eps}}

 \vspace{2mm}

 \leavevmode
 \epsfysize=6cm
 \centerline{\epsfbox{fig6b.eps}}
 \caption{For the case $\det M_\chi>0$, TEDM ($d\rightarrow s\gamma$)
    is  plotted as a function of the mass $m_{\chi^2}$ of the
    heavier chargino for various
    values of $A$ in the case (a) $M_2+\mu>0$ and (b) $M_2+\mu<0$.
    We have chosen parameters: universal
    scalar mass $m=100$ GeV, $\tan \beta =10$, and the mass of
    the light chargino $m_{\chi^1}=50$ GeV.}
 \label{tedmresult}
\end{figure}

\end{document}